\documentclass[twoside,twocolumn,9pt
]{article}
\usepackage{extsizes}
\usepackage[super,sort&compress,comma]{natbib} 
\usepackage[version=3]{mhchem}
\usepackage[left=1.5cm, right=1.5cm, top=1.785cm, bottom=2.0cm]{geometry}
\usepackage{balance}
\usepackage{mathptmx}
\usepackage{sectsty}
\usepackage{graphicx} 
\usepackage{lastpage}
\usepackage[format=plain,justification=justified,singlelinecheck=false,font={stretch=1.125,small,sf},labelfont=bf,labelsep=space]{caption}
\usepackage{float}
\usepackage{fancyhdr}
\usepackage{fnpos}
\usepackage[english]{babel}
\addto{\captionsenglish}{%
  
}
\usepackage{array}
\usepackage{droidsans}
\usepackage{charter}
\usepackage[T1]{fontenc}
\usepackage[usenames,dvipsnames]{xcolor}
\usepackage{setspace}
\usepackage[compact]{titlesec}
\usepackage{hyperref}
\usepackage{epstopdf}%This line makes .eps figures into .pdf - please comment out if not required.

\definecolor{cream}{RGB}{222,217,201}

\begin{document}

\pagestyle{fancy}
\thispagestyle{plain}
\fancypagestyle{plain}{
%%%HEADER%%%
\renewcommand{\headrulewidth}{0pt}
}
%%%END OF HEADER%%%

%%%PAGE SETUP - Please do not change any commands within this section%%%
\makeFNbottom
\makeatletter
\renewcommand\LARGE{\@setfontsize\LARGE{15pt}{17}}
\renewcommand\Large{\@setfontsize\Large{12pt}{14}}
\renewcommand\large{\@setfontsize\large{10pt}{12}}
\renewcommand\footnotesize{\@setfontsize\footnotesize{7pt}{10}}
\makeatother

\renewcommand{\thefootnote}{\fnsymbol{footnote}}
\renewcommand\footnoterule{\vspace*{1pt}% 
\color{cream}\hrule width 3.5in height 0.4pt \color{black}\vspace*{5pt}} 
\setcounter{secnumdepth}{5}

\makeatletter 
\renewcommand\@biblabel[1]{#1}            
\renewcommand\@makefntext[1]% 
{\noindent\makebox[0pt][r]{\@thefnmark\,}#1}
\makeatother 
\renewcommand{\figurename}{\small{Fig.}~}
\sectionfont{\sffamily\Large}
\subsectionfont{\normalsize}
\subsubsectionfont{\bf}
\setstretch{1.125} %In particular, please do not alter this line.
\setlength{\skip\footins}{0.8cm}
\setlength{\footnotesep}{0.25cm}
\setlength{\jot}{10pt}
\titlespacing*{\section}{0pt}{4pt}{4pt}
\titlespacing*{\subsection}{0pt}{15pt}{1pt}
%%%END OF PAGE SETUP%%%

%%%FOOTER%%%
\fancyfoot{}
\fancyfoot[LO,RE]{\vspace{-7.1pt}\includegraphics[height=9pt]{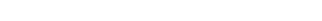}}
\fancyfoot[CO]{\vspace{-7.1pt}\hspace{13.2cm}\includegraphics{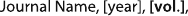}}
\fancyfoot[CE]{\vspace{-7.2pt}\hspace{-14.2cm}\includegraphics{head_foot/RF}}
\fancyfoot[RO]{\footnotesize{\sffamily{1--\pageref{LastPage} ~\textbar  \hspace{2pt}\thepage}}}
\fancyfoot[LE]{\footnotesize{\sffamily{\thepage~\textbar\hspace{3.45cm} 1--\pageref{LastPage}}}}
\fancyhead{}
\renewcommand{\headrulewidth}{0pt} 
\renewcommand{\footrulewidth}{0pt}
\setlength{\arrayrulewidth}{1pt}
\setlength{\columnsep}{6.5mm}
\setlength\bibsep{1pt}
%%%END OF FOOTER%%%

%%%FIGURE SETUP - please do not change any commands within this section%%%
\makeatletter 
\newlength{\figrulesep} 
\setlength{\figrulesep}{0.5\textfloatsep} 

\newcommand{\topfigrule}{\vspace*{-1pt}% 
\noindent{\color{cream}\rule[-\figrulesep]{\columnwidth}{1.5pt}} }

\newcommand{\botfigrule}{\vspace*{-2pt}% 
\noindent{\color{cream}\rule[\figrulesep]{\columnwidth}{1.5pt}} }

\newcommand{\dblfigrule}{\vspace*{-1pt}% 
\noindent{\color{cream}\rule[-\figrulesep]{\textwidth}{1.5pt}} }

\makeatother
%%%END OF FIGURE SETUP%%%

%%%TITLE, AUTHORS AND ABSTRACT%%%
\twocolumn[
  \begin{@twocolumnfalse}
{\includegraphics[height=30pt]{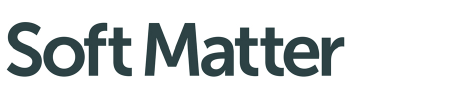}\hfill\raisebox{0pt}[0pt][0pt]{\includegraphics[height=55pt]{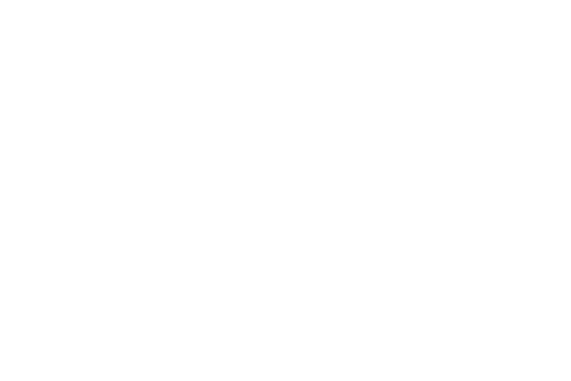}}\\[1ex]
\includegraphics[width=18.5cm]{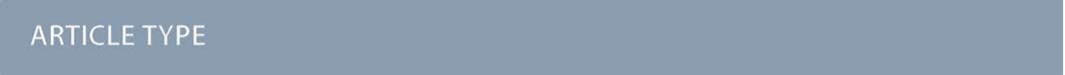}}\par
\vspace{1em}
\sffamily
\begin{tabular}{m{4.5cm} p{13.5cm} }

\includegraphics{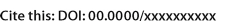} & \noindent\LARGE{\textbf{Repulsion and attraction in the interactions of opposite membrane deformations}} \\%Article title goes here instead of the text "This is the title"
\vspace{0.3cm} & \vspace{0.3cm} \\

 & \noindent\large{Ali Azadbakht,\textit{$^{a}$} and Daniela J. Kraft $^{\ast}$\textit{$^{a}$}} \\%Author names go here instead of "Full name", etc.

\includegraphics{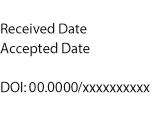} & \noindent\normalsize{
%Membrane proteins cooperate to regulate many cellular functions. This cooperation involves complex forces that act indirectly across membranes between proteins. One of the most important indirect forces stems from membrane deformation. 
Lipid membrane deformations have been predicted to lead to indirect forces between the objects that induce these deformations. Recent experimental measurements have found an attractive interaction between spherical particles that all induce a deformation towards the inside of a giant unilamellar vesicle.
Here, we complement these experimental observations by investigating the interactions between deformations pointing in opposite directions with respect to the membrane normal vector. This is experimentally realized by a particle deforming the membrane towards the inside of the GUV and pulling a membrane tube towards the outside of the membrane. Particles completely wrapped by the membrane are repelled from the tube with a strength of 3~k$_B$T at a distance of 0.5~$\mu$m. However, particles that strongly curve the membrane by adhering only to a patch of about 50~\% of its surface area are attracted to the center of the tube with a strength of -5.3~k$_B$T at a minimum distance of about 1~$\mu$m. We find that such Janus particles also experience attractive interactions when both deforming the membrane in the same way. These quantitative experimental observations provide new insights into interactions between oppositely membrane deforming objects, important for cooperative protein assembly at or interactions of microplastics with cell membranes.} \\%The abstrast goes here instead of the text "The abstract should be..."

\end{tabular}

 \end{@twocolumnfalse} \vspace{0.6cm}

  ]

\renewcommand*\rmdefault{bch}\normalfont\upshape
\rmfamily
\section*{}
\vspace{-1cm}

\footnotetext{\textit{$^{a}$~Soft Matter Physics, Huygens-Kamerlingh Onnes Laboratory, Leiden University, PO Box 9504, 2300 RA Leiden, the Netherlands. kraft@physics.leidenuniv.nl}}

\section{Introduction}

Membrane proteins constitute one-third of all human proteins~\cite{Overington2006a}, which control cellular functions such as receptor signaling and ion transport~\cite{Almen2009}. Errors in their organization are linked to various diseases~\cite{Meng,Suzuki2006,Lee2021}, making the study of the driving forces for protein arrangements in the plasma membrane crucial. Complex proteins migrate in the fluid membrane and interact through many forces, such as electrostatic~\cite{Zhang2011}, capillary~\cite{Kralchevsky1994}, Van der Waals~\cite{Roth1996}, hydrophobic and hydrophilic forces~\cite{Killian1998}, depending on their structure, orientation, electrical charge, and distance to other proteins. 

Besides these well-known interactions, the membrane itself also can induce forces between them~\cite{Dan1993,Goulian1993,Podgornika1996,Reynwar2008}. First, membrane fluctuations can induce interactions through Casimir-like forces, which has been shown to  promote Shiga toxin aggregation~\cite{Pezeshkian2017}. In addition, membrane deformations for example induced by proteins on the membrane~\cite{Goulian1993,Weikl1998} can induce a second type of interaction stemming from the minimization of membrane bending energy and tension. The latter is more difficult to identify and quantify in a biological setting due to the multitude of forces involved. Therefore, simplified models are employed that allows selective and quantitative measurements of these interactions. 

To make theoretical predictions for the interactions between membrane-deforming objects, membrane inclusions with a simplified shape, such as a cone or sphere, are typically considered. Early theoretical considerations found that conical inclusions in a flat membrane repel each other, both for equally and oppositely oriented deformations of the membrane. ~\cite{Goulian1993,Fournier1997,Kim1998}.
%with a power of $s^{-4}$ with the distance $s$ between inclusions~\cite{Goulian1993,Kim1998}. The interaction furthermore depended on the square of the object's contact angle to the membrane, implying that stronger deformation can enhance the interaction strength~\cite{Fournier1997}. Strikingly, the relative orientation of the two inclusions with respect to the membrane, though, was predicted to not be important: both equally oriented and oppositely oriented deformations were predicted to repel. 

Later analytical approaches supported repulsions between objects on the same side, but found an attraction between objects on opposite sides of the membrane by various methods, such as analytical approaches to describe the geometry of the deformed membrane~\cite{Weikl2003,Muller2005a,Muller2007,Muller2010,Mkrtchyan2010} or effective field theory~\cite{Yolcu2012}. 
Numerical simulations that included membrane tension supported this and observed equally oriented inclusions to repel at all distances and oppositely membrane-deforming objects to attract each other at intermediate distances and strongly repel at short ranges. close.~\cite{Weikl1998,Weitz2013}. 

In contrast, coarse-grained simulations and numerical predictions for strongly membrane deforming objects observed a switch from repulsion to attraction when the deformations were oriented in the same direction~\cite{Reynwar2007,Reynwar2011a,Bahrami2018,Zhu2022,Zhu2023_softmatter}. These predictions were in line with recent experimental measurements on simplified model systems consisting of membrane-deforming colloidal spheres and giant unilamellar vesicles. For equally deforming particles, an attraction was consistently found~\cite{Koltover1999,VanDerWel2016,Sarfati2016,Azadbakht2023BPG}, although the strength of the attraction varied considerably, ranging from a few k$_B$T to hundreds of k$_B$T. This variation highlights that minute differences in the induces curvature can play a crucial role in the interaction between membrane-deforming objects.

No experimental measurements exist, though, for interactions between objects that induce opposite curvature on a membrane. While theoretical and numerical predictions for interactions between opposite membrane inclusions have been put forward, the same studies did not show agreement with experiments for equal membrane deformations, suggesting that they may not be able to predict the interactions correctly. 

In this paper, we therefore use a previously developed simplified model system composed of a Giant Unilamellar Vesicle (GUV), which is made of a lipid bilayer and adhesive colloidal particles that deform the membrane to study the interactions between oppositely membrane-deforming colloids. We use both fully adhesive as well as partially adhesive (Janus) colloidal particles to examine the force between equal and opposite membrane deformations. Opposite membrane deformations are realized by pulling a membrane tube from a GUV and studying the interaction with a particle that deforms the GUV towards the inside. We find that fully wrapped particles that induce small deformations are repelled from the pulled tube, whereas partially wrapped particles which create large deformations are attracted. 
%We quantify the interaction of spheres inducing a negative interaction with a pulled tube from the GUV that deforms the membrane with positive curvature. 
To complete the picture, we directly measure the attraction force between two partially wrapped particles deforming a membrane and found it to be attractive in line with earlier measurements on fully-wrapped particles.

\section{Materials and Methods}

\subsection{Chemicals} 
Phosphate-buffered saline (PBS) tablets,
chloroform (99\%),
sodium phosphate (99\%),
D-glucose (99\%),
4,4'-Azobis(4-cyanovaleric acid) (98\%, ACVA),
N-Hydroxysulfosuccinimide sodium salt (98\%), 
Bovine Serum Albumin (BSA),
and 1,3,5,7-tetramethyl-8-phenyl-4,4-difluoro-bora-diaza-indacene (97\%, BODIPY\textsuperscript{TM} FL)
were purchased from Sigma-Aldrich; 
methoxypoly(ethylene) glycol amine (mPEG, M\textsubscript{W}= 5000) from Alfa Aesar;
1-Ethyl-3-(3-dimethylaminopropyl) carbodiimid hydrochloride (99\%, EDC) were obtained from Carl Roth;
NeutrAvidin (avidin) from Thermo Scientific;
1,2-dioleoyl-sn-glycero-3-phosphoethanolamine-N-[biotinyl(polyethylene glycol)-2000] (DOPE-PEG-biotin),
1,2-dioleoyl-sn-glycero-3-phosphocholine (DOPC),
1,2-dioleoyl-sn-glycero-3-phosphoethanolamine-N-(lissamine rhodamine B sulfonyl) (DOPE-Rhodamine),
from Avanti Polar Lipids were purchased from Sigma-Aldrich.
Deionized water with resistivity of 18.2 $M\Omega\cdot c$m obtained from a Millipore Filtration System (Milli-Q Gradient A10) was used in all experiments. All chemical were used as received.

\subsection{Vesicle production}
Giant Unilamellar Vesicles were prepared with a mixture of 97.5 wt.\% DOPC, 2.0 wt.\% DOPE-PEG2000-Biotin, and 0.5 wt.\% DOPE-Rhodamine by the electroformation method as described in \cite{Azadbakht2023nano,Azadbakht2023BPG}.

\subsection{Particle Functionalization}
Carboxylated polystyrene (PS) particles with diameter of 0.98 $\pm$ 0.03 $\mu$m were prepared by surfactant-free radical polymerization \cite{Appel2013}. The fluorescent dye BODIPY was included in the synthesis for imaging. PS particles were functionalized with NeutrAvidin and mPEG 5000 following the procedure described in \cite{VanDerWel2016,VanDerWel2017a}.

\subsection{Sample preparation}
25~mm round coverslips were coated with 0.5~$g/l$ BSA buffer for 2~min and then washed three times with PBS buffer. The BSA-coated coverslips allowed the particles to stick to the coverslip such that GUVs could partially attach to the substrate or particles. Vesicles were gently washed in isotonic PBS and then diluted to remove small lipid aggregates and, then mixed with the particles in the same buffer. Finally, the mixture was injected into a home-built stainless steel microscope chamber. The chamber was kept open for approximately 30 minutes to allow the osmolality of the external solution to increase. All experiments performed in the room temperature. 

\subsection{Microscopy}
Images were captured on an inverted Ti-E Nikon microscope equipped with an A1-R confocal scanner. Excitation light was passed through a 60$\times$ water immersion objective (N.A.=1.2) and the emitted light reflected back through the same light path to the detectors. Particles containing BODIPY and GUVs labeled with Rhodamine were simultaneously excited with 488 and 561~nm laser beams, respectively. The emitted light was collected in the ranges of 500-550~nm (particles, BODIPY) and 580-630~nm (vesicles, rhodamine). Fig.\ref{fig:Fig.1}d corresponds to channel 1\&2 showing fluorescence from membrane and colloid, respectively. Both laser beams scanned a 512*256 pixel field of view with a frame rate of 59 fps in resonant mode.

\subsection{Optical Trapping and Force Measurement}
An optical trap was provided by a highly focused laser beam with a 1064 nm Nd:YAG laser from LaserQuantum beam integrated into the confocal light path to simultaneously image and trap. The laser beam front was modulated by a high-speed Meadowlark spatial light modulator (SLM) with a refresh rate of 120 Hz, and holograms were generated at 100 Hz using RedTweezers software \cite{Bowman2014a}.

The optical tweezers were calibrated using the equipartition theorem and $1/2k_{OT} \langle\Delta x^2\rangle=1/2 k_BT$, where $\Delta x$ is the displacement of a trapped particle from the center of the trap and $k_{OT}$ is the trap stiffness. As long as the displacement of the trapped particle from the center of the trap is small, Hooke's law holds so that the force ($F$) can easily be calculated by $F=k_{OT}\Delta x$.

\subsection{Image analysis and Particle tracking}
The open-source Python package Trackpy\cite{Allan2021Soft-matter/trackpy:V0.5.0} was used for locating the particle center, and circletracking\cite{vanderWel2016CircletrackingV1.0} was employed to find the center and radii of GUVs by fitting an ellipse to its contour.
We define our coordinate system such that the origin 
($x$,$y$,$z$)=(0,0,0) is located at the center of the GUV.
Three-dimensional tracking of particles was achieved by assuming that particles are confined to an ellipsoidal membrane, and the height of particles was derived with 
$z=b\sqrt{1-\frac{x^2}{a^2}-\frac{y^2}{b^2}}$
where $x$,$y$, and $z$ are relative coordinates of particles with respect to the GUV, $a$ is the major radius, and $b$ is the minor radius of the ellipsoid. By combining the particle coordinates with the vesicle geometry, the three-dimensional positions of the particles relative to the vesicle center were determined.

\section{Results and Discussion}
\begin{figure*}
\centering
\includegraphics[width=1\linewidth]{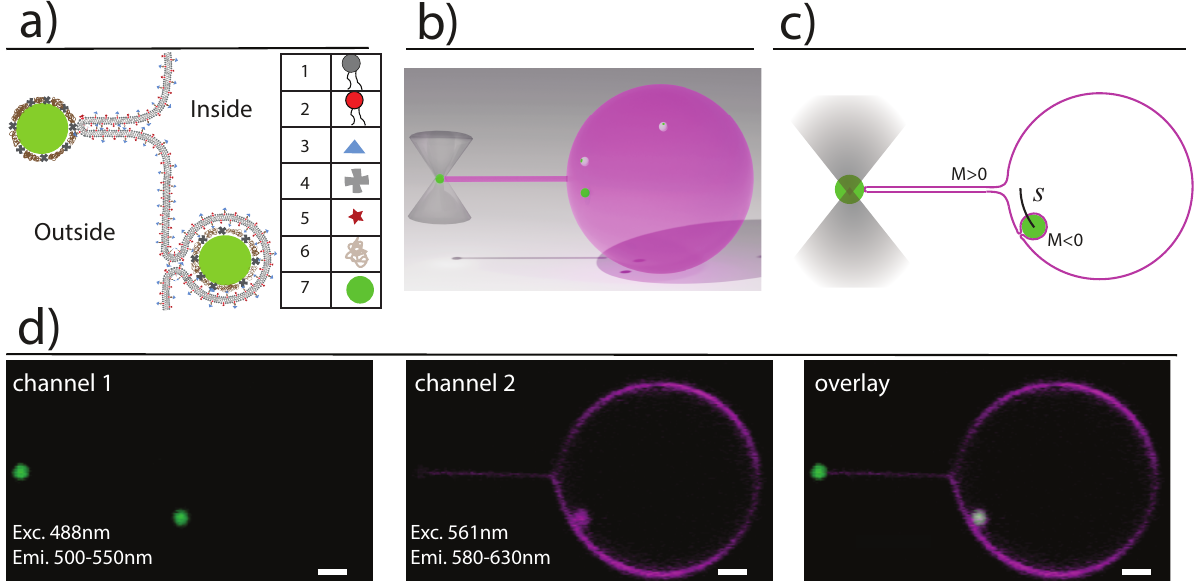}
\caption{\textbf{Experimental setup for measuring the inverse deformation-mediated interaction} 
a) Detailed schematic of the pulled membrane tube creating a deformation to the outside and a wrapped particle deforming membrane toward the inside (not to scale). 1-DOPC lipid 2-DOPE lipid 3-Biotin 4-NeutrAvidin 5-Rhodamine, 6-polyethylene glycol (PEG) 7-Polystyrene particle. 
b) Schematic of the three-dimensional setup of a  GUV (magenta) with a tube pulled with an optical trap and including wrapped, partially-wrapped and non-wrapped particles (green) (not to scale)
c) Schematic of a cross-section of a GUV highlighting positive and negative curvatures (M) and the geodesic distance from the tubes.
d) Confocal images of GUV cross sections in two separated channels, from left to right: (channel 1) particle emission 500-550nm, (channel 2) membrane emission 565-625nm (Overlay);  scale bars are 2~$\mu$m;
}
\label{fig:Fig.1}
\end{figure*}

%\subsection{GUVs and colloidal particles as a model system}
To measure the curvature-mediated interaction of two inclusions, we use a model system based on a Giant Unilamellar Vesicle (GUV) composed of a lipid bilayer and adhesive spherical polystyrene particles. As shown in Fig.\ref{fig:Fig.1}a, adhesion between the colloids and GUV is achieved by NeutrAvidin proteins attached to the outer surface of the colloids which exhibit strong affinity for the biotinylated lipids which are integrated in the GUV \cite{VanDerWel2017}. Each NeutrAvidin-biotin bonds is 17~k$_B$T strong and can be considered permanent on the experimental time scale.

The functionalized particles are wrapped by the GUV when the adhesion energy surpasses the energy required for bending the membrane around the particle if membrane tension is sufficiently low \cite{VanDerWel2016,Spanke2020,Azadbakht2023BPG}. By controlling the NeutrAvidin concentration on the colloids we regulate the adhesion energy and thus the wrapping.\cite{VanDerWel2017a} Wrapping was induced by keeping the microscope chamber open for about 30~min such that water could evaporate from phosphate buffered saline (PBS) media. Accordingly, the membrane tension continuously decreased due to the increasing osmolarity difference until adhesion dominated and the particles were wrapped, see methods for more details.

%current experimental setup and using optical tweezers to pull the tube

%To create a system imposing two opposite signs of curvature, similar to previous studies\cite{VanDerWel2016,Azadbakht2023}, two wrapped particles were required, 
One way to create an experimental system with two deformations with opposite curvature would be to use two wrapped particles: one particle inside the GUV that is being wrapped towards the outside, and a second particle that is wrapped from the outside towards the inside.
However, while standard encapsulation methods such as inverted emulsions can create GUVs with particles on the inside,~\cite{Moga2019} the thus prepared GUVs possess spontaneous curvature.~\cite{Morita2021}  Our attempts to use this method therefore resulted in the spontaneous formation of tubes that extended both inward and outward from the membrane, similar to ref. \cite{Dasgupta2018}, making measurements prohibitively difficult.

For this reason, we took an alternative approach to obtain deformations with opposite curvature. In addition to using a particle wrapped from the outside to the inside as before, we created a membrane deformation with opposite curvature by pulling a particle attached to the outside of the GUV further outwards using optical tweezers. This induces the formation of a tube and in turn a positive curvature deformation in the membrane at the point where the tube was connected to the membrane. See Fig.\ref{fig:Fig.1}b,c,d for the experimental setup. 

From the force $F$ that is applied to pull the tube and the bending rigidity $\kappa$, we can also extract the membrane tension $\sigma$, using $\sigma=\frac{F^2}{8\pi^2\kappa}$ \cite{Simunovic2016}. See Methods section for force measurement details. We quantified the membrane tension by assuming $\kappa$ = 22~k$_{B}$T for DOPC vesicles created by electroformation~\cite{Rawicz2000,Faizi2020} and found $\sigma$  to be in a range from 3.6 to 6.8 $\mu$N/m. 
The tube diameter then was estimated from  $d_{tube}=\sqrt{\frac{2\kappa}{\sigma}}$ to be in the range of 160 to 220 nm. ~\cite{Simunovic2016}  

To image how the wrapped particle interacts with the GUV, we fluorescently labeled the membrane with 0.5 wt\% Rhodamine-lipids and the particles with BODIPY dye. Fast confocal microscopy in combination with a python-based image analysis routine~\cite{Allan2021Soft-matter/trackpy:V0.5.0} then allows both easy tracking of particles and characterization of the GUV, such as their size and the position of their center~\cite{vanderWel2016CircletrackingV1.0}(Fig.\ref{fig:Fig.1}d ). We monitored the colocalization of the fluorescence signal in the channels to distinguish whether the particles were wrapped or not. This can be seen in Fig.\ref{fig:Fig.1}d , where a white color in the overlay of GUV (magenta) and colloid (green) indicates membrane-wrapped particles.

The membrane tube was drawn from the equatorial plane of the GUV (Fig.\ref{fig:Fig.1}b, c, d) where the focal plane of the confocal microscope was located. Taking the equator as $z=0$, the z-position of the wrapped particles was extracted from knowledge of the x-y position of the particle and the constraint that it had to be confined to the membrane. To do so, we fitted the membrane in the vicinity of the tube-induced deformation with an ellipsoidal shape (see Materials and Methods for details).

For symmetry reasons, the interaction energy between the two opposite inclusions is only related to their geodesic distance $s$, which is shown in Fig.\ref{fig:Fig.1}c, and here $s$ is the distance between the wrapped particle and the intersection of the tube and GUV. To quantify the interaction, various techniques have been developed to extract the free energy from the trajectory of the particles \cite{Gnesotto2020,Frishman2020,Merrill2009,Sarfati2017,Crocker1994} Since the wrapped particles were not in the focal plane of the confocal microscope most of the time, we could not use techniques that require a long and/or continuous trajectory.\cite{Gnesotto2020,Frishman2020}. 
Therefore, we employed a displacement-based energy calculation relying on a master equation here.\cite{Crocker1996} Assuming the diffusion coefficient to be constant, we calculated the transition probability matrix directly from the probability that a particle moves from $s_i$ to $s_j$. The interaction energy $u(s)$ between opposite inclusions is then simply determined from the Boltzmann distribution\cite{Crocker1994,VanDerWel2016}.

\subsection{Fully wrapped particles repel from a pulled tube}

\begin{figure*}
\centering
\includegraphics[width=1\linewidth]{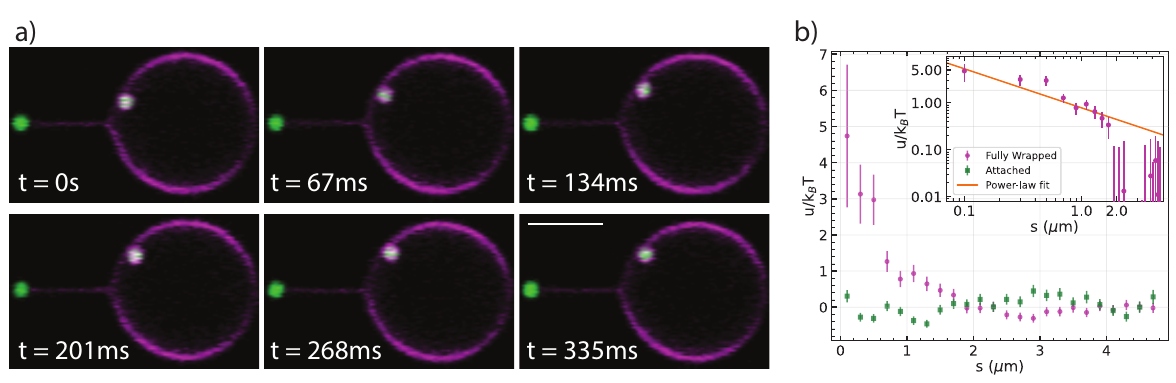}
\caption{\textbf{Interaction between a pulled tube inducing a positive deformation of the GUV and a wrapped particle inducing a negative deformation of the GUV} a) Confocal microscopy time series showing the motion of a particle wrapped towards the inside of the GUV and a tube being pulled by a particle towards the outside of the GUV. Scale bar is 5~$\mu$m. See also SI movie S1. 
b) Interaction energy as a function of geodesic distance $s$ between the tube and a fully-wrapped and an attached particle depicted by magenta circles and green squares, respectively. Inset: a power-law fit (red line) with $u/u_0=\alpha s^\beta$ to the interaction between the fully-wrapped particles and the tube yields $\alpha=0.78\pm 0.14$ and $\beta =-0.83\pm0.08$. 
}
\label{fig:Fig.2}
\end{figure*}

We first measured how a particle that was fully wrapped by the membrane interacted with the oppositely curved deformation induced by the tube. In contrast to previous work where two identical inclusions were found to attract,\cite{VanDerWel2016,Sarfati2016} the wrapped particle here was repelled from the opposite curvature induced by the tube, as can be seen in the image sequences in Fig.\ref{fig:Fig.2}a and Supplementary Video S1. %The Supplementary Video S1 shows wrapped particles and how they interact with the tube. While the wrapped particle is repelled, the motion of the unwrapped particle that does not distort the membrane does not seem to be affected by the presence of the tube. 

We extract the interaction energy from the particle's trajectory using the transition probability matrix approach and show the result in Fig.\ref{fig:Fig.2}b. We find that the interaction is indeed repulsive and that the energy decays with distance up to about 2~$\mu$m. The decay of the repulsive energy follows a power-law in the distance with an exponent of -0.80 $\pm$ 0.08, see inset of Fig \ref{fig:Fig.2}b. At longer distances, the potential energy reaches a plateau which we assume corresponds to the energy in the absence of any membrane deformation and therefore the average of the energy at 4 to 5 $\mu$m distance is set to zero. Our observation is very different from early predictions assuming a flat and tensionless membrane as well as small membrane deformations, which did not find a different sign in the interaction between opposite and equal membrane inclusions, and a power law decay of $s^{-4}$.\cite{Goulian1993} However, none of the underlying assumptions hold in our case.

Comparing the strength of the repulsion with the previously observed attraction between two equal inclusions, we find that although the sign of the interaction has inverted, the strength is of similar magnitude. 
For two equal inclusions, the minimum in the attraction was found to be -3.3$k_B$T which occurred at a distance of about 1.3~$\mu$m.~\cite{VanDerWel2016} It was not possible to measure distances closer than one particle diameter, equivalent to $s=0.98 \mu$m, for steric reasons. Here, the wrapped particle can approach the tube much closer, as the tube is taking up significantly less space than another wrapped particle. 

At very close distances of the particle from the membrane neck, we find a repulsion of almost 5~$k_B$T. A repulsion similar in strength as the attraction for two equal inclusions, i.e. 3~$k_B$T,  is found at  $s=0.5 \mu$m, which is closer than the distance where this interaction strength occurs for two equal inclusions(Fig. \ref{fig:Fig.2}b). The interaction decays over about 2.5 $\mu$m for the equal and 2.0$\mu$m distance for opposite inclusions. The here observed slightly shorter interaction range and lower strength at the same distance likely arise from a higher membrane tension, which is in line with measurements on equal inclusions at higher membrane tensions.\cite{VanDerWel2016} In addition, the shape of the deformation induced by the tube may also differ from that induced by a fully but oppositely wrapped particle which would affect the interaction energy.

To test whether the observed interaction was only due to the curvature imposed by the particle we also measured the interaction of a particle with the tube that was only adhered to but did not deform the membrane. To ensure that the particles did not leave an indentation in the membrane, the control experiment was performed with particles whose surface density of NeutrAvidin was reduced by a factor of four, such that only 2.5\% of their surface was covered with NeutrAvidin.~\cite{VanDerWel2016}  We find that there is no measurable interaction between the tube and the non-deforming particles as depicted in Fig.\ref{fig:Fig.2}b (green squares). This implies that the deformation induced by the tube does not severely affect the geometry of the membrane surrounding it. It also demonstrates that interactions mediated by fluctuation are negligible in this system \cite{Agudo-canalejo2017,Li2017a}.

The repulsive interaction found in this experiment resembles the membrane-mediated interaction of two spheres on an elongated vesicle.~\cite{Vahid2017} These simulations showed that the membrane-deforming particles stayed far away from the positively curved regions most of the time and were pushed away from the stretching points~\cite{Vahid2017}, in agreement with our observation that wrapped particles were repelled from a tube with positive curvature. 
% Numerical solutions found that the power-law factor of the potential energy as a function of distance was also reduced from -5 for weak deformation to -1 for strong deformation both on a same side of the membrane.\cite{Reynwar2011a} The current results suggest analogous trends for deformations of opposite signs. Strong curvatures could affect the potential form nonlinearly and convert an attractive potential into a repulsive potential.
 
%ellipsoid membranes that tube region fitted with the parabola

\subsection{Partially-wrapped particles attract to a pulled tube}
Incidentally, partially wrapped particles were found to emerge when they were in the vicinity of the tube during the wrapping transition. 
Although the engulfment process regularly takes no more than a few seconds, if the membrane does not have enough available surface area, the particle remains half-covered by the membrane for some time.

\begin{figure*}
\centering
\includegraphics[width=1\linewidth]{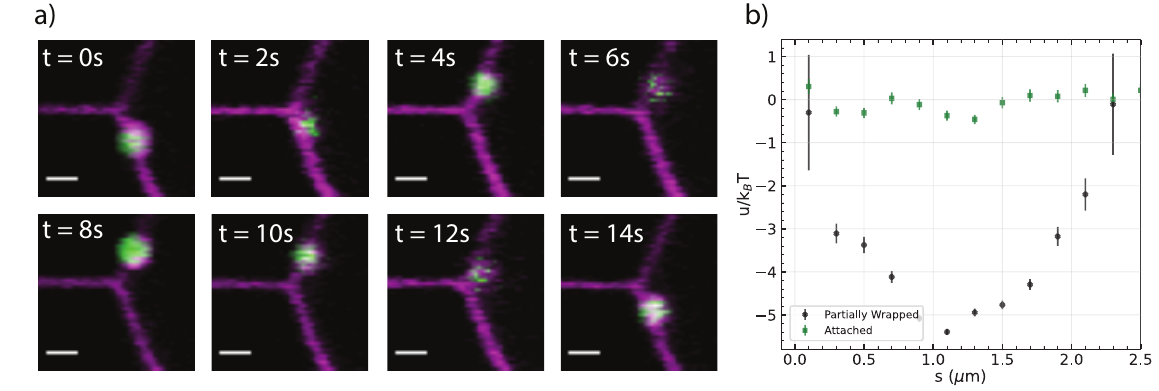}
\caption{\textbf{Interaction between a pulled tube and partially-wrapped particle} a) Time evolution of the wrapped particle from left to right and top to bottom; the time interval between snapshots 2s (scale bar 1$\mu$m)
b) Interaction energy as a function of geodesic distance ($s$) between the center of the tube and a partially-wrapped (black crosses) and attached, non-wrapped particle (green squares).}
\label{fig:Fig.3}
\end{figure*}

In Figure \ref{fig:Fig.3}a, a time sequence of a partially-wrapped particle in the vicinity of the tube is shown. These particles became fully wrapped after the tube was released due to the decreasing membrane tension and excess membrane area. The attraction between a partially-wrapped particle and the outward tube is obvious as it is confined to positions near the tube (Supporting Video S2).

The same method as in the previous section is applied to evaluate the potential energy. As shown in Figure~\ref{fig:Fig.3}b, the partially wrapped particle moves in an attractive  potential with a depth of -5.3~ k$_B$T, which is located about 1~$\mu$ m from the center of the tube. The shape of the potential well is parabolic by approximation. The particle senses the tube from a distance of 2~$\mu$m and becomes attracted. Assuming a Hookian force and using a parabolic fit we found a potential well with a stiffness of 51.5 $\pm$ 0.5 fN$/\mu$m. Here, we directly measured the membrane tension from the force applied to the membrane tube, yielding a value of $\sigma$ = 4.3 $\pm$ 1.8 $\mu$N/m.

% Confocal microscopy does not have a sufficient spatial resolution to accurately determine how deeply the particle is wrapped in the membrane. 
To measure the wrapping fraction of the particle, we estimated the distance of the particle center to the undisturbed membrane to be 70~$\pm$~78~$n$m by repeatedly measuring it using confocal microscopy images from the GUV's equator (see Figure S1 in Supporting Information). This corresponds to 53$\pm$4\% of the particle being covered by the membrane.

In comparison with previous analytical and simulation studies, we find similarities and differences. Most of the predictions found attractions between two objects deforming the membrane from opposite sides \cite{Weikl2003,Muller2010,Muller2007,Mkrtchyan2010,Yan2019a}. Within the particular range of membrane tensions used in our experiment a similar potential well has been calculated which is comparable to the interaction energy we find here in Figure~\ref{fig:Fig.3}b \cite{Weikl1998, Weitz2013}. However, that minimum of the potential well was found to be on the order of the membrane bending rigidity ($\kappa$ = 22 k$_B$T), in contrast to the depth of our almost parabolic potential well at -5.3 k$_B$T.

Moreover, it is striking that we find a repulsion from the opposite deformation of the tube for fully wrapped particles and an attraction for partially wrapped particles. A similar transition from repulsion to attraction was observed by coarse-grained molecular dynamics simulations and numerical calculations for axisymmetric particles in a flat membrane.\cite{Reynwar2011} Upon an increasing contact angle, i.e. an increasing curvature imprint, the interaction switched from a repulsion to an attraction at short distances and repulsion at longer distances. While we don't see a repulsion at larger distances, we here find a similar switch when for the larger deformations induced by partially wrapped particles.

\subsection{Attraction between adhesive Janus particles}

\begin{figure*}
\centering
\includegraphics[width=0.8\linewidth]{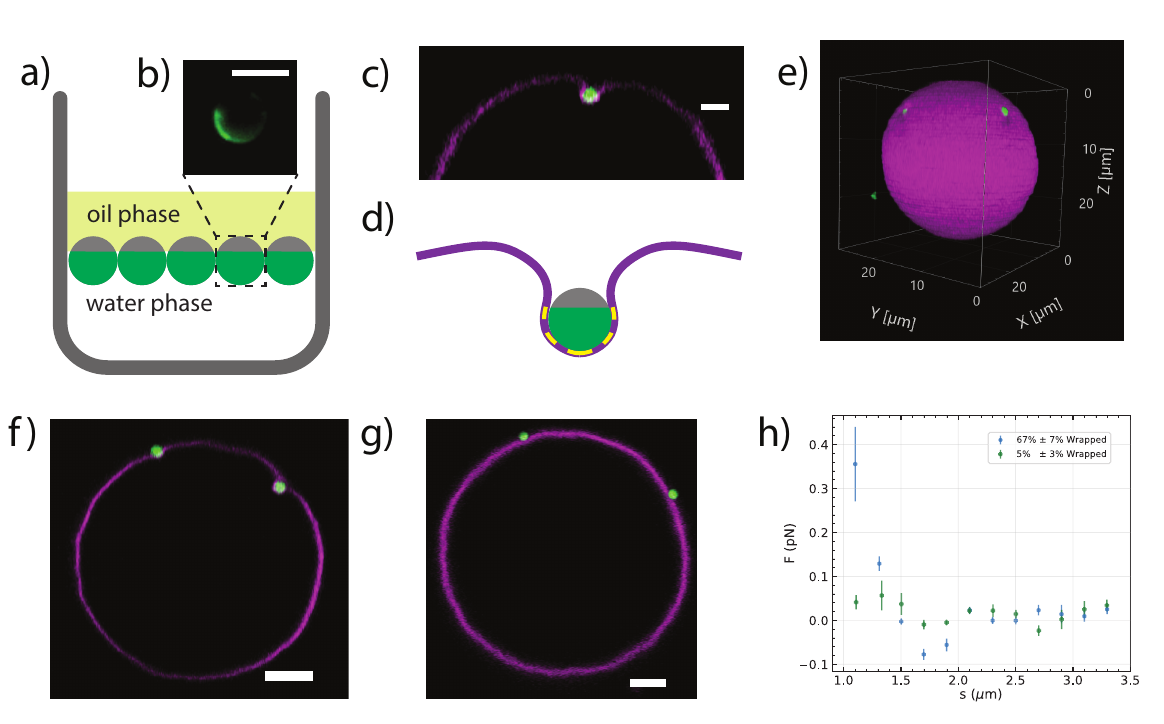}
\caption{\textbf{Interaction between two partially-wrapped particles} 
a) Schematic representation of the functionalization approach used to prepare colloids partially coated with NeutrAvidin (not to scale); particles were adsorbed to an oil-water interface and selectively functionalized on the area exposed to the water phase.
b) Confocal image of a non-fluorescent polystyrene coated with NeutrAvidin conjugated with fluorescein used to validate the method clearly shows a fluorescent cap, indicating the Janus character of the particle;
c) Confocal image and d) schematic representation showing that a partially wrapped colloid (green) strongly deforms the membrane (magenta); scale bar is 2~$\mu$m;
e) 3D confocal reconstruction of a GUV with two particles embedded (particle shown in green, vesicle in magenta);
f) A confocal image of a GUV (magenta) with two adhered Janus particles (green) with path size equal to 67 $\pm$ 7\% of its surface area; 
g) Confocal microscopy image of a GUV (magenta) with two adhered Janus particles (green) with patch size equal to 5 $\pm$ 3\% of its surface area; scale bar are 5~$\mu$m;
h) The interaction force exerted on each Janus particle measured as a function of the distance between their centers $s$. Blue hexagonal data points represent interactions between two colloids with 67\% wrapping fraction and green plus-shaped data points represent two Janus colloids with 5\% wrapping fraction.}
\label{fig:Fig.4}
\end{figure*}

To have better control over the membrane deformation, we devised a protocol in which the polystyrene spheres were intentionally only partially functionalized with NeutrAvidin, so called Janus particles. To do so, the particles were adsorbed to an interface between oil and water and were functionalized by NeutrAvidin on the surface that was exposed to the water phase, see Figure \ref{fig:Fig.4}a. The functionalized fraction of the particles' surface area is equal to the fraction of the particles' surfaces exposed to water. The coating fraction can then be estimated from the contact angle of the particles at the interface, which were 75$^\circ$\cite{Al-Shehri2014} for particles absorbed to a dodecane-water interface corresponding to 63\% of the area, and 22$^\circ$ for absorption to an octanol-water interface corresponding to only 7\% of the area exposed to Neutravidin~\cite{Sabapathy2015VisualizationNanoparticles}. 

It is assumed that the functionalized fraction is equal to the adhered area due to the strong adhesion energy of 17 $k_BT$ between NeutrAvidin and biotin \cite{Moy1994}. We validated our coating mechanism using a non-fluorescent particle which we coated with NeutrAvidin conjugated with a fluorescein dye and observed by confocal microscopy, shown in Figure \ref{fig:Fig.4}b. To quantify the precise adhesion area, we determined the contact length between the Janus particle and the vesicle with the confocal microscope and divided it by the circumference of the particle, see Figure~\ref{fig:Fig.4}c and the yellow dashed line indicated in the schematic of Figure~\ref{fig:Fig.4}d. In the cases where the membrane has been strongly deformed, the particle overshoots when moving into the vesicle during wrapping and the distance of the particle from the undisturbed membrane is more than one particle diameter, which can thus not be used to simply measure the wrapping fraction area~\cite{Dietrich1997,Agudo-canalejo2017,Spanke2020,Azadbakht2023nano, ewins_controlled_2022}. These strong deformations are clearly visible in the 3D reconstruction of the membrane, as shown in Figure~\ref{fig:Fig.4}e, and different from earlier work on penetration depth of Janus particles.\cite{ewins_controlled_2022} We find an adhesive fraction of particles area of 67\% $\pm$ 7\% for particles (Figure~\ref{fig:Fig.4}f) functionalized at the dodecane-water interface and 5\% $\pm$ 3\% for particles at the octanol-water interface (Figure~\ref{fig:Fig.4}g).

We furthermore observed how two Janus particles which strongly deform the membrane aggregate and find that the attraction is so strong that the pair once formed cannot be broken by thermal fluctuation during the experimental time framework, see Supporting Video S3. The strong attraction precludes the use of statistical methods such as the Boltzmann weighing or the transition probability matrix that has been used in the first two sections, as we cannot obtain sufficient statistics for all canonical microstates in our experimental time frame. Therefore, we here instead used two optical traps to measure the force between the particles directly~\cite{Azadbakht2024ACS}. We chose to trap the particles at the top of the vesicle and hold one particle in a strong trap while the other particle was moved with respect to this trap acting as a force sensor, see Supporting Video S4. We then measured the force ($F$) at each relative distance between the centers of the two particles ($s$), see Figure \ref{fig:Fig.4}h. The attractive force increases the closer the particles are to each other and has a maximum of 0.4~$pN$ when the particles are almost touching. Interestingly, there is a small repulsive force between the Janus particles at 1.7 $\mu$m distance. An important factor in membrane deformations is the bendocapillary length ($\lambda_{bc}=\sqrt{\frac{\sigma}{\kappa}}$) that $\kappa$ = 22~k$_{B}$T for DOPC vesicles\cite{Faizi2020}. The membrane tension $\sigma$ was measured from the fluctuation spectrum of the vesicles in the equatorial contour and was found to be in the range of less than 10 nN/m \cite{Pecreaux2004}. Such a low membrane tension increases the bendocapillary length $\lambda_{bc} \geq$ 3$ \mu$m, implying that length scales smaller than $\lambda_{bc}$ are dominated by bending energy~\cite{Deserno2004}. To verify that the measured interaction is due only to the deformation and that the employed optical traps do not perturb the vesicle, we quantify the interaction between two Janus particles with small wrapping fraction and find that they do not interact significantly interact, see Figure~\ref{fig:Fig.4}g and h.

Our force measurement for two Janus particles deforming the membrane on the same side shows an attraction, similar to previous experiments on completely wrapped spheres. The interaction between two fully wrapped particles was previously found to be reversible with 0.1~$p$N~\cite{VanDerWel2016}. Here, we Measure a significantly larger, non-reversible interaction, possibly due to the strong deformations of the partially wrapped particles, see Figure~S2. Another experiment with larger adhesive colloids yielded an attractive force of about 1~$p$N at close contact~\cite{Sarfati2016}, which is in line with our experiment where a stronger deformation is induced by control over the adhesion area of the particles. 

Our results for two Janus particles are also very similar to Monte Carlo and molecular dynamics simulations~\cite{Bahrami2018,Zhu2022}. In these simulations a strong attraction at close distances and a local repulsion minimum around 3$R_p$ was found as well as a small repulsion peak at larger distances. However, in our experiments, two Janus particles do not measurably interact with each other at distances larger than 2~$\mu$m, while in the simulations the repulsive force continued to increase slightly, probably due to the larger relative size of the colloids to the vesicles.

\section{Conclusion}
We complemented the experimental evidence of how two membrane deforming particles interaction with experiments on partially wrapped particles that induce deformations on the same or opposite side of the membrane. We quantified the membrane-mediated interaction stemming from the deformation of two inclusions with opposite curvature, where one was experimentally realized by a pulled membrane tube. We found repulsion between opposite inclusions when small deformations as created by fully membrane-wrapped particles were employed. However, for stronger deformations of the membrane which were achieved by employing partially wrapped particles, we find attractions to the opposite deformation of the membrane tube in line with predictions by Reynwar and Deserno.\cite{Reynwar2011} 
Control experiment with non-wrapped particles proved that the interaction was solely due to the curvature. To better understand the effects of membrane deformation induced by partially wrapped particles, we created Janus colloidal spheres which only possessed an attractive patch. In agreement with previous simulations, we measured an attractive force between two spheres, which had  67\% of their surface area adhered to the membrane. We then quantified this interaction force using two optical traps and found it to be about 0.4~$p$N at near touching contact distance.

Our current model system holds the potential to advance our understanding of biological cell membranes. As biological membranes encompass multiple components and distinct liquid phases~\cite{Veatch2003}, each with its own mechanical properties~\cite{Needham1988}, our model provides a platform to quantitatively explore the interaction energy between two membrane deforming objects on such membranes further. Moreover, the controlled manipulation of local curvature, whether positive or negative, could potentially initiate dynamic shifts within a specific phase, prompting interesting avenues for future investigation.

\section*{Acknowledgment}
We gratefully acknowledge useful discussions with Thomas Weikl, and Markus Deserno. We would like to thank Yogesh Shelke for his assistance in demonstrating the technique for placing particles at the interface of oil and water.

\section*{Author Contributions}
A.A. and D.J.K. designed, performed the research and analyzed the data and wrote the the paper.

\section*{Conflicts of interest}
There are no conflicts to declare.

% \section*{Data Availability Statement}
% The data supporting this study, including an Excel file with separate sheets containing the data required to generate Figures 1b, 2b, and 3h, is available on 4TU.ResearchData at the following link: \href{https://data.4tu.nl/private_datasets/oBdRfoLAVCfo_A32k56IQF952SduHP5H3T_S8b7ayx8}{Link} and will be released upon acceptance of the manuscript.

\balance

\bibliography{references_1}
\bibliographystyle{rsc} 

\end{document}